\documentclass[10pt]{article}
\usepackage{graphicx}
\usepackage{lineno}

\def\Title#1{\begin{center} {\Large #1 } \end{center}}
\def\Author#1{\begin{center}{ \sc #1} \end{center}}
\def\Address#1{\begin{center}{ \it #1} \end{center}}

\newcommand\pubblock{\rightline{\begin{tabular}{l} Proceedings of the Second Annual LHCP\\ \pubnumber\\
         \pubdate  \end{tabular}}}

\newenvironment{Abstract}{\begin{quotation} \begin{center} 
             \large \end{center}\bigskip 
      \begin{center}\begin{large}}{\end{large}\end{center} \end{quotation}}

\newenvironment{Presented}{\begin{quotation} \begin{center} 
             PRESENTED AT\end{center}\bigskip 
      \begin{center}\begin{large}}{\end{large}\end{center} \end{quotation}}

\def\beq{\begin{equation}}
\def\eeq#1{\label{#1}\end{equation}}
\def\eeqn{\end{equation}}

\def\beqa{\begin{eqnarray}}
\def\eeqa#1{\label{#1}\end{eqnarray}}
\def\eeqan{\end{eqnarray}}

\let\bar=\overbar

\def\Dslash{\not{\hbox{\kern-4pt $D$}}}
\def\dslash{\not{\hbox{\kern-2pt $\del$}}}

\def\msb{{\bar{\ssstyle M \kern -1pt S}}}

\usepackage{color}
\RequirePackage{xspace}
\def\Dbar {\kern 0.2em\overline{\kern -0.2em D}{}\xspace}
\def\DzDstarzb{\ensuremath{D^0\kern -0.16em \Dbar^{*0}}\xspace}

\newcommand\pubnumber{ ATL-PHYS-PROC-2014-121 }

\newcommand\pubdate{\today}

\def\support{\footnote{Current address: Pacific Northwest National Laboratory, Richland, WA 99352, U.S.A. }}

\def\affiliation{
On behalf of the ATLAS, CMS, LHCb, and ALICE Experiments \\
SLAC National Accelerator Laboratory \\
Menlo Park, CA 94025, U.S.A.
\support }

\begin{document}

\large
\begin{titlepage}
\pubblock

\vfill
\Title{QUARKONIA PRODUCTION AND POLARIZATION AT THE HADRON COLLIDERS}
\vfill

\Author{ Bryan FULSOM  }
\Address{\affiliation}
\vfill
\begin{Abstract}

This talk presents a review of recent results for quarkonium production at the LHC from ATLAS, CMS, LHCb, and ALICE.
Production cross sections for $J/\psi$, $\psi(2S)$, and $\Upsilon(mS)$, and production ratios for $\chi_{c,bJ}$ are found to be in good agreement with predictions from non-relativistic QCD.
In contrast, spin-alignment (polarization) measurements seem to disagree with all theoretical predictions.
Some other production channels useful for investigating quarkonium hadroproduction mechanisms are also considered.

\end{Abstract}
\vfill

\begin{Presented}
The Second Annual Conference\\
 on Large Hadron Collider Physics \\
Columbia University, New York, U.S.A \\ 
June 2-7, 2014
\end{Presented}
\vfill
\end{titlepage}
\def\thefootnote{\fnsymbol{footnote}}
\setcounter{footnote}{0}
%

\normalsize 


\section{Introduction}
Quarkonium is the bound state of heavy quarks ($c\overline{c}$ and $b\overline{b}$).
It is a generally well-understood system, with many successful predictions for masses, widths, decay modes, and production rates for an entire family of particles.
The study of quarkonium production at the hadron colliders is of particular interest, not only because it is an important test of QCD, but also because there are several emerging conflicts between experimental results and theoretical predictions.
This talk covers recent results related to the production cross sections and spin alignment (polarization) of quarkonia at the LHC.

The two leading models to describe quarkonia hadroproduction are the color singlet model (CSM) and non-relativistic QCD (NRQCD) (also referred to as the color octet model, or COM, in the presentation).
In the CSM, the heavy quark pair is produced in a color singlet state and evolves into the final state quarkonium with the same quantum numbers.
In NRQCD, color octet states can also be produced, which then evolve to the singlet final state via soft gluon emission.
Matrix elements for the various color octet contributions are determined from a fit to the data.
NRQCD \cite{ref:bodwin} was originally proposed in order to explain shortcomings of the CSM in predicting production rates in early CDF data \cite{ref:braaten, ref:cdf}.
The NRQCD calculations resulted in excellent agreement with the production rate data, though this success is partially expected by design.
Higher order corrections to the CSM have also provided improved agreement with experimental results seen in the Tevatron experiments \cite{ref:artoisenet}.

A detailed discussion of the techniques for measuring quarkonium polarization is provided by \cite{ref:faccioli}.
For spin-1 quarkonia ($J/\psi$, $\psi(2S)$, and $\Upsilon(1S,2S,3S)$), the spin alignment of the produced quarkonium is quantified by the relative importance of the spin-1 eigenstates, determined by measuring the angular distribution of the leptonic pair decay (\emph{e.g.} $J/\psi\rightarrow\mu^{+}\mu^{-}$).
The most general description of the system requires two angles ($\theta$ and $\phi$) measured relative to an arbitrary quantization axis, and three polarization parameters ($\lambda_{\theta}$, $\lambda_{\phi}$, and $\lambda_{\theta\phi}$).
The CSM and NRQCD have considerably different theoretical predictions for the spin alignment, making this observable an important discriminant between these two leading quarkonium hadroproduction theories.
Unfortunately, past results from the Tevatron experiments disagree with each other at a level greater than ${\sim}4\sigma$, and appear to support neither production model \cite{ref:faccioli, ref:cdf_new}.

This review of recent quarkonia production results consists of three parts: production cross sections, polarization measurements, and other tests of production mechanisms.

\section{Production Cross Sections}
All quarkonia hadroproduction measurements at the LHC generally follow a similar analysis strategy.
The spin-1 quarkonia are reconstructed via their decays to two muons (due to triggering/background, decays to electrons are not typically considered), requiring an accurate understanding of detector muon acceptance.
Signal events are determined from the $m_{\mu^{+}\mu^{-}}$ distribution, with backgrounds extrapolated from the sidebands.
$\chi_{c,bJ}$ quarkonia are reconstructed via radiative decays (\emph{e.g.}: $\chi_{cJ}\rightarrow\gamma J/\psi$).
Photon energy resolution sufficient to distinguish between the different $J$ states is achieved by identifying photons converting into $e^{+}e^{-}$ pairs in detector material.
Non-prompt quarkonia (\emph{i.e.}: $B\rightarrow c\overline{c} X$) are separated from prompt production by vertex-related variables taking advantage of the measureable flight distance of the secondary decay from the primary vertex.
Measurements of the production rate are then performed in bins of, for example, $p_{T}$, $|y|$, cos$\theta$, and $\phi$.

Production cross section measurements have been performed by the four major LHC experiments.
In essentially all cases, the NRQCD predictions are a better fit to the data than those from the CSM or others, though the CSM with next-to-next-to-leading order corrections \cite{ref:artoisenet} remains competitive.
Recent measurements of the $J/\psi$ production cross section agree with one another and theoretical predictions \cite{ref:jpsi_alice, ref:jpsi_lhcb, ref:psi_cms}.
Measurements of $\psi(2S)$ production \cite{ref:jpsi_alice, ref:psi_cms, ref:psi_lhcb}, including a preliminary result by ATLAS reconstructing $\psi(2S)\rightarrow\pi^{+}\pi^{-}J/\psi$ rather than $J/\psi\to\mu^{+}\mu^{-}$ presented here for the first time \cite{ref:psi_atlas}, also agree with one another and theoretical predictions up to the highest measured values of $p_{T}$ for both prompt and non-prompt production.
However, although the experimental data agree well across experiments, at the highest $p_{T}$ values there is evidence for a departure from the theoretical predictions for prompt, and particularly non-prompt, production.
This disagreement is a subject for future understanding.

Upsilon production cross sections show similar behaviour.
LHCb \cite{ref:ups_lhcb} and ALICE \cite{ref:jpsi_alice} have covered the low-$p_{T}$ region (up to 15 GeV/$c$), while CMS \cite{ref:ups_cms} and ATLAS \cite{ref:ups_atlas} extend up to $p_{T}<50$ GeV/$c$ and $p_{T}<70$ GeV/$c$, respectively.
Theoretical predictions show reasonable agreement with experimental measurements that gradually worsen with increasing $p_{T}$.
Part of this disagreement could be due to the fact that feed-down effects (\emph{e.g.} contributions to the final state $\Upsilon(pS)$ decays from $\Upsilon(mS)$ and $\chi_{b}(nP)\to X\Upsilon(nS)$) are not accounted for in the theoretical predictions, but are present in the experimental data.

For the P-wave states, ATLAS \cite{ref:chic_atlas}, CMS \cite{ref:chic_cms}, and LHCb \cite{ref:chic_lhcb} have performed measurements of $\chi_{cJ}(1P)$ production.
The ATLAS result is the most recent, and the first at the LHC to measure absolute production rates.
This requires a detailed knowledge of the conversion photon characteristics in the ATLAS detector.
In all cases, the results are well-described by the NRQCD predictions.
To date, the only measurement in the bottomonium sector is a preliminary result by CMS \cite{ref:chib_cms}.
Analagous to $\chi_{cJ}$, radiative decays $\chi_{bJ}(1P)\to\gamma\Upsilon(1S)$ are reconstructed and use the improved resolution of converted photons to separate the $J=1,2$ peaks.
This measurement is more difficult than for $\chi_{cJ}(1P)$ due to the smaller ${\sim} 20$ MeV energy splitting, and the CMS result is the first at the LHC to be able to resolve the $\chi_{b1,2}$ peaks.

\section{Spin Alignment}
\begin{figure}[htb]
\centering
\includegraphics[height=4.5in]{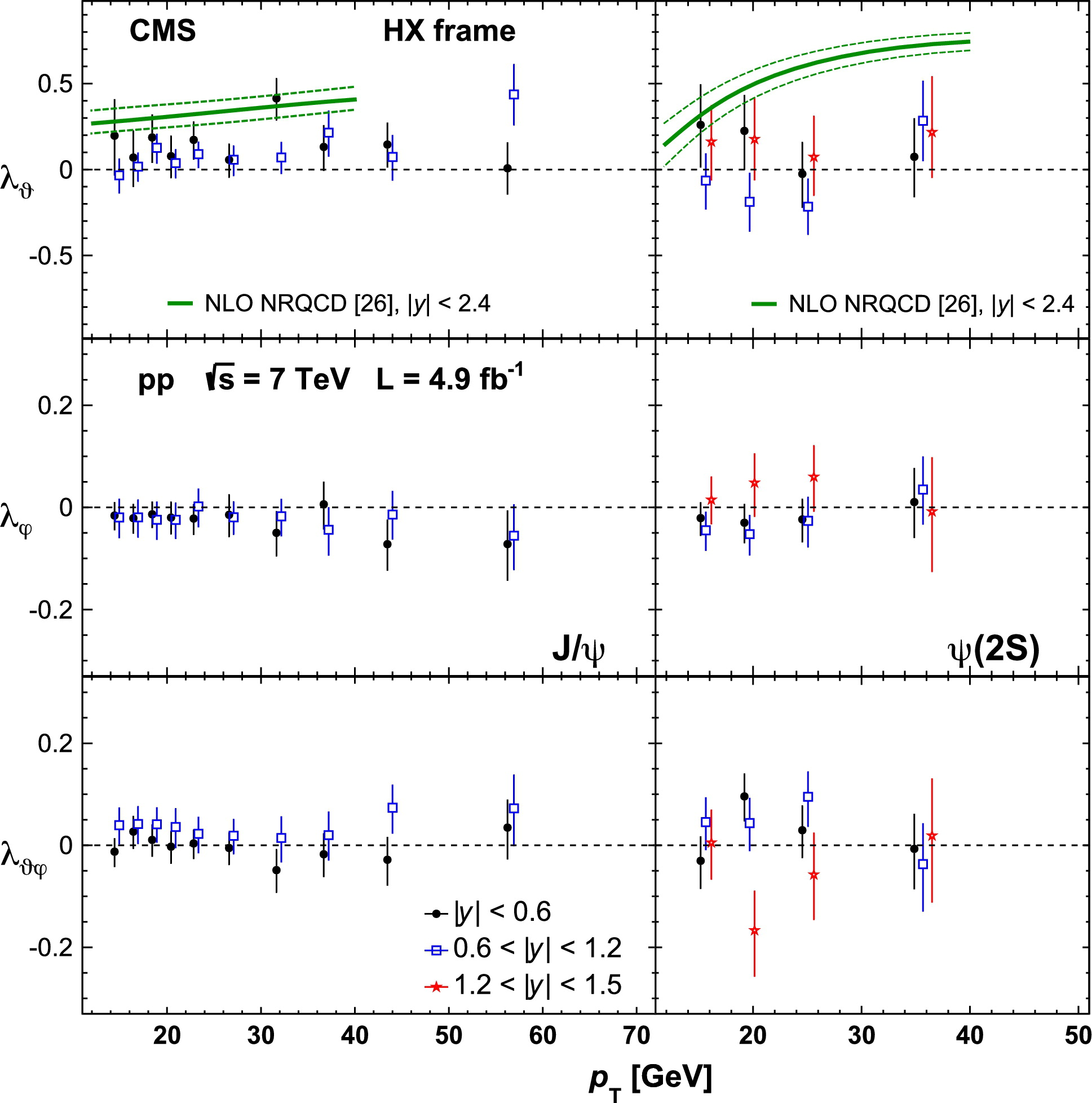}
\caption{From \cite{ref:jpsipolar_cms}: CMS measurement of polarization parameters versus $p_{T}$ for $J/\psi$ (left) and $\psi(2S)$ (right) in different bins of rapidity and frames of reference. The theoretical predictions (listed as Reference [26] in the figure above) are from \cite{ref:cmsplot_theory}. The results are consistent with no polarization, and disagree with the theoretical predictions.}
\label{fig:cms}
\end{figure}

ALICE, LHCb, and CMS have published results on quarkonia spin alignment (polarization) \cite{ref:alice_previous, ref:jpsipolar_lhcb, ref:jpsipolar_cms}, with the latter two results (LHCb and CMS) being the most recent and therefore highlighted in detail here.
These two experiments offer a complementary range of $p_{T}$ and rapidity coverage.
Both analyses closely followed the prescription set forth in \cite{ref:faccioli}: all three polarization parameters are measured in multiple frames of reference.
For $J/\psi$ production, LHCb \cite{ref:jpsipolar_lhcb} and CMS \cite{ref:jpsipolar_cms} find very little to no polarization.
Neither the CSM nor the NRQCD predictions of direct $J/\psi$ agree with these experimental results.
Because these measurements are of inclusive $J/\psi$, there is a possibility that $\psi(2S)$ and $\chi_{cJ}(1P)$ feed-down effects may be ``washing out'' polarization effects.
$\psi(2S)$ decays should be free from this possible source of contamination.
However, neither LHCb \cite{ref:psipolar_lhcb} nor CMS \cite{ref:jpsipolar_cms} observe polarization in this system either.
Figure \ref{fig:cms}, taken from CMS \cite{ref:jpsipolar_cms}, shows the measured polarization parameters versus $p_{T}$ for $J/\psi$ (left) and $\psi(2S)$ (right) in different bins of rapidity and frames of reference.
The predictions from a recent theoretical calculation \cite{ref:cmsplot_theory} are shown in green; the disagreement is clearly evident.

Considering bottomonium, $\Upsilon(1S,2S,3S)$, again shows similar results.
CMS analysed the production of all three states and find small or no polarization \cite{ref:upspolar_cms}.
It should be noted that this system is not free from feed-down effects from higher quarkonium ($\Upsilon(nS)$ and $\chi_{bJ}(nP)$ states ranging up to $n=3$).
Nonetheless, measureable polarization effects were predicted by all theoretical models, and are conspicuously absent for all types of quarkonia production and in all frames of reference considered at the LHC thus far.
This represents a new challenge for theorists going forward.

\section{Other Tests of Quarkonia Production}
This section covers three selected recent results that test other aspects of quarkonia production predictions.

A charmonium-like state known as the $X(3872)$ was discovered by the Belle experiment about 10 years ago \cite{ref:x3872_belle}.
It is now widely believed to be a four-quark state, most likely a charm meson (\DzDstarzb) molecule, or perhaps an exotic tetraquark state.
Based on results from the Tevatron, a theoretical non-relativistic quantumchromodynamical $X(3872)$ production rate prediction exists \cite{ref:x3872_theory}.
CMS reconstucted $X(3872)\to\pi^{+}\pi^{-} J/\psi$ decays to measure prompt $X(3872)$ production \cite{ref:x3872_cms}.
The cross section results versus $p_{T}$ show that the theory predicts the correct $p_{T}$ dependence but is too low by approximately an order of magnitude.
This is indicative that further work is needed on the theoretical side related to the contributions considered in the calculation, or perhaps with the understanding of the $X(3872)$ as a \DzDstarzb bound state.

In a recent preliminary result, CMS has studied the production of double prompt $J/\psi$ events \cite{ref:dbljpsi_cms}.
Based on a fit to invariant $\mu^{+}\mu^{-}$ masses, decay length, and separation distance significance, $446\pm23$ prompt double $J/\psi$ events originating from the same vertex were found.
The results show no evidence of possible $\eta_{b}(1S)\to J/\psi J/\psi$ decays, but based on an enhancement in $|\delta y|$ between the $J/\psi$ candidates, potentially hint at double parton scattering production of $J/\psi$.

Finally, ATLAS has published the first evidence for prompt $W + J/\psi$ production \cite{ref:jpsiw_atlas}.
Theoretically, color singlet production is expected to dominate color octet, making this channel a potential discriminator between production models.
The ATLAS analysis reconstructed $J/\psi\to\mu^{+}\mu^{-}$ and $W\to\mu\nu$, and performed a fit to the invariant mass of the lepton pair and the pseudo-proper time.
The contribution from double parton scattering was also estimated and subtracted, resulting in a prompt $W + J/\psi$ production rate about an order of magnitude larger than expected.

\section{Conclusions}
In summary, all of the LHC experiments have now provided important quarkonium production results, spanning a wide range of $|y|$ and $p_{T}$.
These results have provided new insights into quarkonium production mechanisms.
In terms of production cross section, NRQCD predictions agree best with the data.
However, perhaps most interesting is that there is no evidence for any polarization, which is in disagreement with the leading theories.

Going forward, most of the experiments have yet to exploit their full Run-I datasets, and will soon add greater statistics and probe a new center-of-mass energy regime with the addition of Run-II data.
They will be able to extend the reach in $p_{T}$ and further explore $\chi_{c}$ and $\chi_{b}$ production.
It is clear that more theoretical work is also needed, particularly to explain the discrepancy in the polarization results.
One such early attempt is to include the LHC results in a fit to the data to guide theoretical models \cite{ref:future_theory}.
Another important avenue of future research is to consider ``quarkonium $+$ X'' channels, where ``X'' can be an object (such as a $W$ or $Z$) less affected by increasing trigger thresholds, for example.
It is clear that the LHC will continue to play a key role in solving the puzzles of quarkonium hadroproduction.





\end{document}